\documentclass[epj]{svjour}
\usepackage{graphicx}
\begin{document}
\title{Pseudospin--Electron Model in Large Dimensions}
\author{I.V. Stasyuk \and A.M. Shvaika}
%
%
\institute{Institute for Condensed Matter Physics, Nat. Acad. Sci. Ukr.,
1~Svientsitskii Str., UA--290011 Lviv, Ukraine}
\date{Received: date / Revised version: date}
%
\abstract{
Energy spectrum and thermodynamics of the pseudospin--electron
model introduced at the consideration of the anharmonicity effects
in high--$T_c$ superconductors are investigated in the dynamical
mean field approximation ($d=\infty$ limit). In the limit of zero
electron correlation $U\to 0$ this model is analytically exactly
soluble within this approach: in the $\mu=\mathrm{const}$ regime
the first order phase transition with the jump of the pseudospin
mean value $\langle S^z\rangle$ and reconstruction of the electron
spectrum can realize, while in the $n=\mathrm{const}$ regime the
phase separation in electron subsystem can take place for certain
values of the model parameters. On the basis of the obtained
results the applicability of the approximate schemes previously
used for the investigation of the pseudospin--electron model are
discussed.
} 
\maketitle

\section{Introduction}\label{intro}

The theoretical investigation of the strongly correlated electron
systems is an enduring subject of interest in condensed matter
physics especially during last ten years after the discovery of
high--$T_c$ superconductivity. Recent studies of strongly
correlated Hubbard type models elucidate some important features of
high--$T_c$, namely the $d$--wave pairing and the role of
antiferromagnetic fluctuations \cite{Dagotto}. However some
features of the cuprates are not well understood (e.g. the
existence of an ``optimal'' doping, the effect of charge
fluctuations, strong electron--phonon interaction, lattice dynamics
and instabilities of ferroelectric type). Thus the extensions of
the Hubbard model by the van Hove scenario concept, the
incorporation of the local electron--phonon interaction (the
Hubbard--Holstein model for harmonic phonons and the
pseudospin--electron model \cite{Muller} for anharmonic ones),
generalization to a two or three band model, the inclusion of
intersite electron interaction, etc. are under consideration.

Within these models the pseudospin--electron one \cite{Muller} in a
simplest way includes the interaction of correlated electrons with
some local lattice excitations described by pseudospins (e.g.
anharmonic vibrations of apex oxygen in YBaCuO type HTSC's), and
shows the possibility of dipole (pseudospin) and charge density
instabilities \cite{PhysC,MolPhysLet} and phase separation
\cite{CzJPh} due to the effective retarded interaction between
pseudospins via conducting electrons. All these results were
obtained within the generalized random field approximation (GRPA)
\cite{Izyumov} which is a realization of the appropriate
perturbation theory for correlation functions in the case of strong
coupling ($U\gg t$) and corresponds to the mean field approximation
in calculation of mean values. There are no good criteria of its
applicability and it is supposed that GRPA gives correct
description in the case of large dimensionality of local (site)
states.

In recent years the essential achievements of the theory of strong
correlated electron systems are connected with the development of
the dynamical mean field theory (DMFT) proposed by Metzner and
Vollhardt \cite{Metzner} for Hubbard model (see also \cite{Kotliar}
and references therein). DMFT is a nonperturbative scheme which
allows to project Hubbard model on the single impurity Anderson
model and is exact in the limit of infinite space dimensions
($d=\infty$). Moreover, some class of models (e.g. Falicov--Kimball
model \cite{FK}) can be studied almost analytically within DMFT.

Here we apply DMFT to the investigation of pseu\-do\-spin--electron
model in the limit of zero electron correlation ($U=0$) which can
be treated analytically.

\section{Perturbation theory in terms of electron transfer}
\label{Sect:1}

The Hamiltonian of pseudospin--electron model in the absence of
electron correlations can be written in the form:
  \begin{equation}
  H=\sum_i H_i + \sum_{ij\sigma} t_{ij} a_{i\sigma}^{\dagger}a_{j\sigma},
  \label{Hamilton}
  \end{equation}
where
  \begin{equation}
  H_i=gS_i^z\sum_{\sigma}n_{i{\sigma}} - \mu\sum_{\sigma}n_{i\sigma}
   - hS_i^z
  \label{1siteH}
  \end{equation}
is single--site Hamiltonian, and includes local interaction of
conducting electrons with pseudospins placed in longitudinal field
$h$ (asymmetry parameter of anharmonic potential).

In general, one--electron Green's function
$G_{\sigma}(\omega_n,\vec{k})$
  \begin{equation}
  G_{ij}^{\sigma}(\tau-\tau')=\left\langle {\cal T} a_{i\sigma}(\tau)
  a_{j\sigma}^{\dagger}(\tau')\sigma(\beta)\right\rangle_0 /
  \left\langle \sigma(\beta)\right\rangle_0
  \end{equation}
  \begin{equation}
  \sigma(\beta)={\cal T}\exp\bigg\{-\int_0^{\beta}\!d\tau
  \sum_{ij\sigma} t_{ij} a_{i\sigma}^{\dagger}(\tau)
  a_{j\sigma}(\tau)\bigg\}
  \end{equation}
satisfies Larkin's equation
  \begin{equation}
  G_{ij}^{\sigma}(\tau-\tau')=\Xi_{ij}^{\sigma}(\tau-\tau')+
  \Xi_{il}^{\sigma}(\tau-\tau'')t_{lm}G_{mj}^{\sigma}(\tau''-\tau'),
  \label{larkin}
  \end{equation}
where summation (integration) over repeated indices is supposed.
The formal solution of eq. (\ref{larkin}) can be written in the
form
  \begin{equation}
  G_{\sigma}(\omega_n,\vec{k})=\frac1{\Xi_{\sigma}^{-1}(\omega_n,\vec{k})-t_\vec{k}}
  \label{GFlat}
  \end{equation}
and the task is to calculate the irreducible according to Larkin
parts $\Xi_{\sigma}(\omega_n,\vec{k})$.

It is convenient to introduce projective operators on pseudospin
states
  \begin{equation}
  P_i^{\pm}=\frac12\pm S_i^z, \quad (P_i^{\pm})^2=P_i^{\pm},
  \quad P_i^+P_i^-=0
  \label{project}
  \end{equation}
and by substitution $P_i^+=c_i$, $P_i^-=1-c_i$ Hamiltonian
(\ref{Hamilton}), (\ref{1siteH}) can be transformed into the
Hamiltonian of binary alloy. On the other hand, if we keep in
(\ref{Hamilton}), (\ref{1siteH}) only electrons with one
orientation of spin by removing the sum over spin indices and
putting $\sigma=\uparrow$ and consider electrons with
$\sigma=\downarrow$ as localised $P_i^+=n_{i\downarrow}$,
$P_i^-=1-n_{i\downarrow}$ we get the Hamiltonian of the
Falicov--Kimball model where $h$ plays a role of the chemical
potential for the localized $\downarrow$--electrons. As a rule the
common chemical potential is introduced for both electron
subsystems but the case of two chemical potentials was also considered (see,
e.g. \cite{Freericks}) and the first consideration of Falicov--Kimball model
within DMFT was done by Brandt and Mielsch \cite{FK}.

Formally diagrammatic series for the irreducible part
$\Xi_{ij}^{\sigma}\left(\omega_n\right)$ are the same for all these
models
  \begin{eqnarray}
  \Xi_{ij}^{\sigma}(\omega_n)=\bigg(
  \raisebox{-9mm}[3mm][9mm]{\includegraphics[height=12mm]{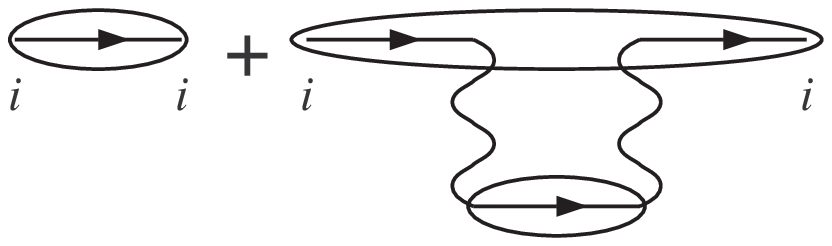}}
  +\ldots\bigg)\delta_{ij}
  \nonumber\\
  +\raisebox{-9mm}[3mm][9mm]{\includegraphics[height=12mm]{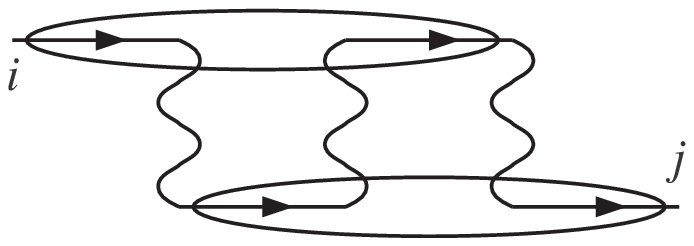}}
  +\ldots
  \label{lambda}
  \end{eqnarray}
and includes both single--site and intersite contributions. Here,
arrows indicate electron propagators $g_{\sigma
i}^{\pm}(\omega_n)=\frac{P_i^{\pm}}{i\omega_n+\mu\mp\frac g2}$ in
the subspaces projected on the pseudospin states of site $i$ and
ovals represent semi--invariant averaging of projection operators.

The main difference between these models is in the way how an
averaging procedure over projection operators is performed (thermal
statistical averaging in the case of pseudospin--electron and
Falicov--Kimball models and configurational averaging for binary
alloy) and how self--consistency is achieved (fixed value of
longitudinal field $h$ for pseudospin--electron model, fixed value
of the component concentration $c$ for binary alloy and fixed value
of the electron concentration --- total or for both electron
subsystems --- for Falicov--Kimball model).

\section{The limit of large dimensions ($d\to\infty$)}
\label{Sect:2}

In the case of high dimensions ($d\to\infty$) one should scale
hopping integral
  \begin{equation}
  t_{ij}\to\frac{t_{ij}}{\sqrt{d}}
  \end{equation}
in order to obtain finite density of states (the Gaussian one for
$d=\infty$ hypercubic lattice $\rho(\varepsilon)=\frac
1{W\sqrt{\pi}} e^{-\varepsilon^2/W^2}$ and semi--elliptic d.o.s.
for $d=\infty$ Bethe lattice $\rho(\varepsilon)=\frac 2{\pi W^2}
\sqrt{W^2-\varepsilon^2}$ \cite{Kotliar}). Due to such scaling only
single--site contributions survive in the expression for
irreducible parts $\Xi_{\sigma}$
  \begin{equation}
  \Xi_{ij}^{\sigma}(\tau-\tau')=\delta_{ij}\Xi_{\sigma}(\tau-\tau'),
  \quad
  \Xi_{\sigma}(\omega_n,\vec{k})=\Xi_{\sigma}(\omega_n)
  \end{equation}
and such site--diagonal function, as it was shown by Brandt and
Mielsch \cite{FK}, can be calculated by mapping the
infinite--dimensional lattice problem on the atomic model
  \begin{eqnarray}
  \lefteqn{e^{-\beta H}\to
  e^{-\beta H_\mathrm{eff}}=
  e^{-\beta H_0}
  }
  \label{atom}\\ \nonumber &&
  \times{\cal T}\!\exp\bigg\{-\int_0^{\beta}\!d\tau \int_0^{\beta}\!d\tau'
  \sum_{\sigma} J_{\sigma}(\tau-\tau') a_{\sigma}^{\dagger}(\tau)
  a_{\sigma}(\tau')
  \bigg\}
  \end{eqnarray}
with auxiliary Kadanoff--Baym field $J_{\sigma}(\tau-\tau')$
\cite{Baym} which has to be self consistently determined from the
condition that the same function $\Xi_{\sigma}$ defines Green's
functions for lattice (\ref{GFlat}) and atomic limit
  \begin{equation}
  G_{\sigma}^{(a)}(\omega_n)=\frac1{\Xi_{\sigma}^{-1}(\omega_n)-J_{\sigma}(\omega_n)}.
  \label{GFatom}
  \end{equation}

``Dynamical'' mean field $J_{\sigma}(\tau-\tau')$ describes the
hopping (transfer) of electron from atom into environment at moment
$\tau$, propagation in environment without stray into atom until
moment $\tau'$. Connection between these ``dynamical'' mean field
of atomic problem and Green's function of the lattice can be
obtained using standard CPA approach \cite{Kotliar}:
  \begin{equation}
  J_{\sigma}(\omega_n)=\Xi_{\sigma}^{-1}(\omega_n) -
  G_{\sigma}^{-1}(\omega_n),
  \end{equation}
where
  \begin{equation}
  G_{\sigma}^{(a)}(\omega_n)=G_{\sigma}(\omega_n)=
  \int_{-\infty}^{+\infty} \!dt \frac{\rho(t)}{\Xi_{\sigma}^{-1}(\omega_n)-t}
  \end{equation}
is a single--site Green's function both for atomic limit and
lattice. Here summation over wave vector was changed by the
integration with the density of states $\rho(t)$.

In order to complete our self--consistent set of equations we
should find expression for Green's function in the atomic limit
(\ref{GFatom}). Due to the properties of the projection operators
(\ref{project}) one can rewrite Hamiltonian of atomic problem
(\ref{atom}) in the form
  \begin{equation}
  e^{-\beta H_\mathrm{eff}} =
  P^+e^{-\beta H^{(+)}} + P^-e^{-\beta H^{(-)}}
  \end{equation}
and our space of states splits into two independent subspaces hence
all projection operators (\ref{project}) act at the same site and
in any order of the perturbation theory expansion all projection
operators can be replaced by their product result and there are no
necessity to make semi--invariant expansions.

Single--electron Green's function is a sum of Green's functions in
subspaces and is equal
  \begin{eqnarray}
  G_{\sigma}^{(a)}(\omega_n) =
  \frac{\left\langle P^+\right\rangle}{i\omega_n+\mu-J_{\sigma}(\omega_n)-\frac g2}
  \\ \nonumber
  +\frac{\left\langle P^-\right\rangle}{i\omega_n+\mu-J_{\sigma}(\omega_n)+\frac g2}.
  \end{eqnarray}

Partition functions in subspaces are
  \begin{eqnarray}\label{partfunc}
  \lefteqn{
  Z_{\pm}=\mathop{\mathrm{Sp}}e^{-\beta H_{\pm}} = e^{\pm\frac{\beta h}2-Q_{\pm}}
  }
  \\ \nonumber
  && =e^{\pm\frac {\beta h}2}
  \prod_{\sigma}\left( 1+ e^{-\beta(\mu\mp\frac g2)}\right)
  \prod_n\left(1-\frac{J_{\sigma}(\omega_n)}{i\omega_n+\mu\mp\frac g2}\right)
  \end{eqnarray}
and presents the partition functions of the non--inter\-ac\-ting
fermions with frequency dependent hopping placed in the external
field formed by pseudospin.

Pseudospin mean value is determined by equation
  \begin{eqnarray}
  \lefteqn{
  \left\langle S^z\right\rangle = \frac12\frac{Z_+-Z_-}{Z_++Z_-}
  }
  \\ \nonumber
  &&=\frac12 \tanh\frac12\left(\beta h-\left( Q_+[\left\langle S^z\right\rangle]-Q_-[\left\langle S^z\right\rangle]\right)\right)
  \label{Sz}
  \end{eqnarray}
which is an analogue of the well known equation of state for Ising
model in mean--field approximation: $\left\langle S^z\right\rangle
=\frac12
\tanh\frac{\beta}2\left( h + J_0\left\langle S^z\right\rangle\right)$.
It should be noted that in the case of Lorentzian density of states
$\rho(\varepsilon)=\frac W{\pi(W^2+\varepsilon^2)}$, which is
frequently used in some applications of DMFT, one can easily obtain
a simple result $J_{\sigma}(\omega_n)=iW$ \cite{Kotliar},
quantities $Q^{\pm}$ do not depend on $\left\langle
S^z\right\rangle$ and equation (\ref{Sz}) transforms into an
expression for $\left\langle S^z\right\rangle$ that indicates the
sensitivity of the equation of state to the shape of d.o.s.

Electron concentration mean value is determined by
  \begin{equation}\label{eq:n}
  \left\langle n\right\rangle=\frac1{\beta}\sum_{m\sigma}G_{\sigma}\left(\omega_m\right)
  \end{equation}
and the functional of thermodynamic potential can be derived in the
same way as it was done in \cite{FK} for Falicov--Kimball model
  \begin{equation}\label{thermpot}
  \frac{\Omega}N = \Omega_{(a)} - \frac1{\beta}\sum_{n\sigma} \bigg\{
  \ln G_{\sigma}^{(a)}(\omega_n) -
  \frac1N\sum_\vec{k}\ln G_{\sigma}(\omega_n,\vec{k})
  \bigg\},
  \end{equation}
where
  \begin{equation}
  \Omega_{(a)}=-\frac1{\beta}\ln(Z_++Z_-)
  \end{equation}
is a thermodynamic potential for atomic problem.

\begin{figure}
  \centerline{
  \includegraphics[width=0.9\columnwidth]{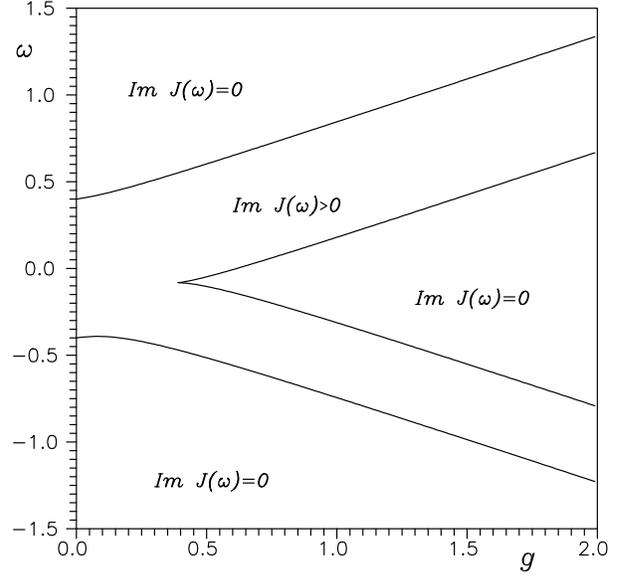}
  }
  \caption{Electron bands boundaries (semi--elliptic d.o.s., $W=0.4$,
  $\left\langle S^z\right\rangle=0.2$).}
  \label{ebands}
  \end{figure}

Below, all calculations will be performed for semi--elliptic
density of states when the auxiliary field is determined by the
simple cubic equation
  \begin{eqnarray}
  \label{eq:J}
  J_{\sigma}(\omega_n)=\frac{W^2}4\left\{
  \frac{\left\langle P^+\right\rangle}{i\omega_n+\mu-J_{\sigma}(\omega_n)-\frac g2}
  \right.
  \\ \nonumber
  +\left.\frac{\left\langle P^-\right\rangle}{i\omega_n+\mu-J_{\sigma}(\omega_n)+\frac g2}\right\}.
  \end{eqnarray}
In a usual way we perform analytical continuation on real axis
($i\omega_n\to\omega-i\delta$) and only solutions of (\ref{eq:J})
with $\Im\mathrm{m}J_{\sigma}(\omega)>0$ must be considered. Band
boundaries are determined from the condition
$\Im\mathrm{m}J_{\sigma}(\omega)\to 0$ and in Fig.~\ref{ebands}
their dependence on coupling constant $g$ are presented.
One can see that there exists critical value of coupling constant
$g\sim W$ when a gap in spectrum appears. It should be noted that
within GRPA as well as in other approaches where single--electron
Green's function is calculated in Hubbard--I approximation, when we
keep only the first term of the single--site contribution in the
expression for the irreducible part (\ref{lambda}), these gap in
spectrum always exists.

In the case of strong coupling ($g\gg W$) an analytical solutions
can be obtained
  \begin{equation}
  J_{\sigma}(\omega)=\frac12\left(\omega\mp\frac g2\right) +
  \frac i2\sqrt{W^2\langle P^{\pm}\rangle-\left(\omega\mp\frac g2\right)^2}
  \end{equation}
for upper and lower subbands, respectively, and one can see that
subbands halfwidth is equal to $W\sqrt{\frac12\pm\langle
S^z\rangle}$ whereas in Hubbard--I approximation it is
$W\left(\frac12\pm\langle S^z\rangle\right)$. This result clearly
shows that even for the case of strong coupling when subbands are
well separated and one of them become narrow ($\left\langle
S^z\right\rangle\to\pm\frac12$) Hubbard--I approximation is
unsufficient and can not be derived from the exact solution in any way, e.g.
due to the subbands halfwidth square root dependence on the localized states
occupance ($\left\langle P^{\mp}\right\rangle\to 0$).

Presented above expressions were obtained for the fixed value of
the chemical potential $\mu$ when stable states are determined from
the minimum of the thermodynamical potential (\ref{thermpot}). This
regime $\mu=\mathrm{const}$ corresponds to the case when the charge
redistribution between conducting sheets CuO$_2$ and other
structural elements (charge reservoir, e.g. nonstoichiometric in
oxygen CuO chains in YBaCuO type structures) which fix the value of
the chemical potential is allowed. On the other hand, in the regime
of the fixed electron concentration value one should solve equation
for chemical potential $n=\langle n\rangle$ (\ref{eq:n}) and stable
states are determined by the minimum of the free energy
$F=\Omega+\mu n$.

\section{Results and discussion}
\label{Sect:3}

\begin{figure}
  \centerline{\includegraphics[width=0.9\columnwidth]{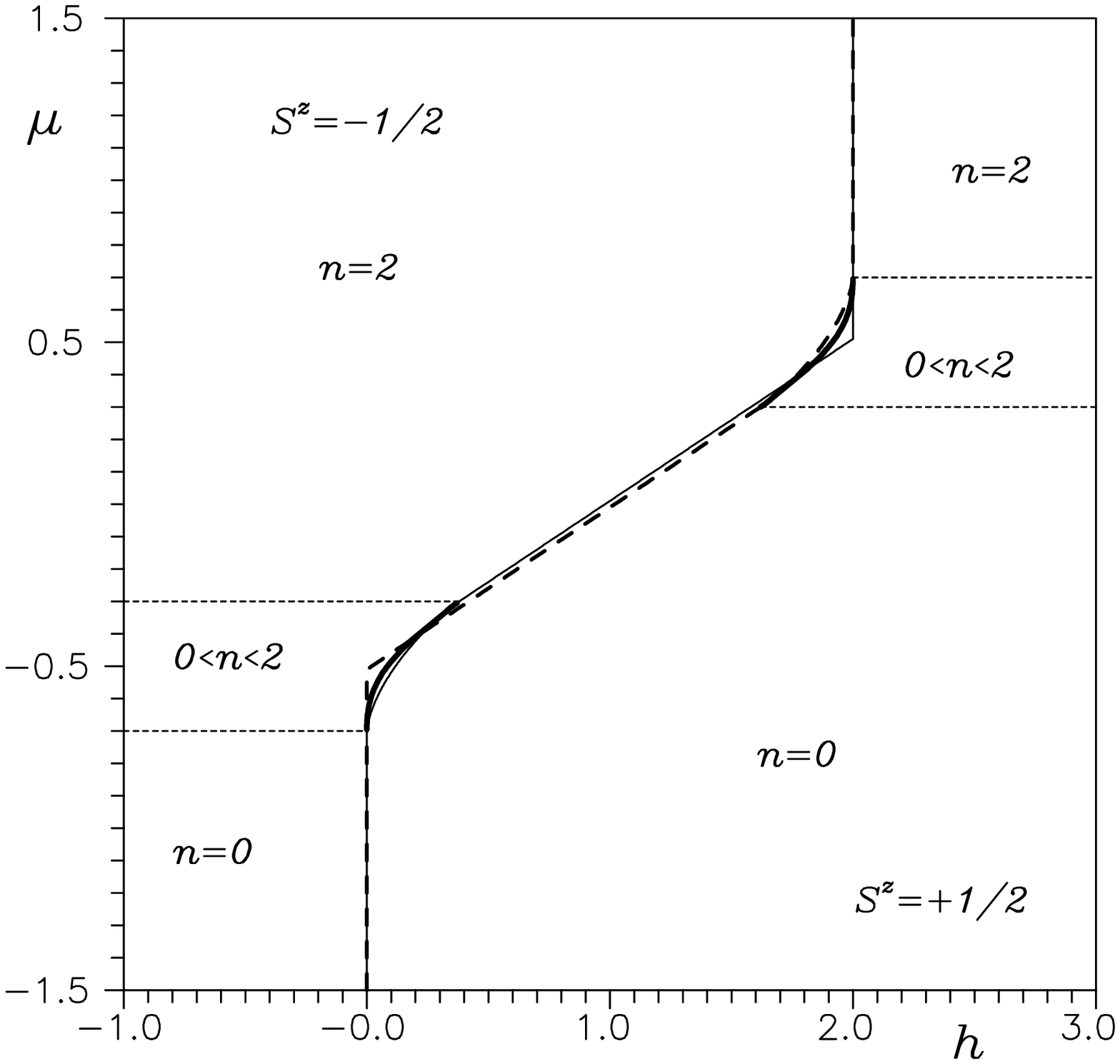}}
  \centerline{a)}
  \bigskip
  \centerline{\includegraphics[width=0.9\columnwidth]{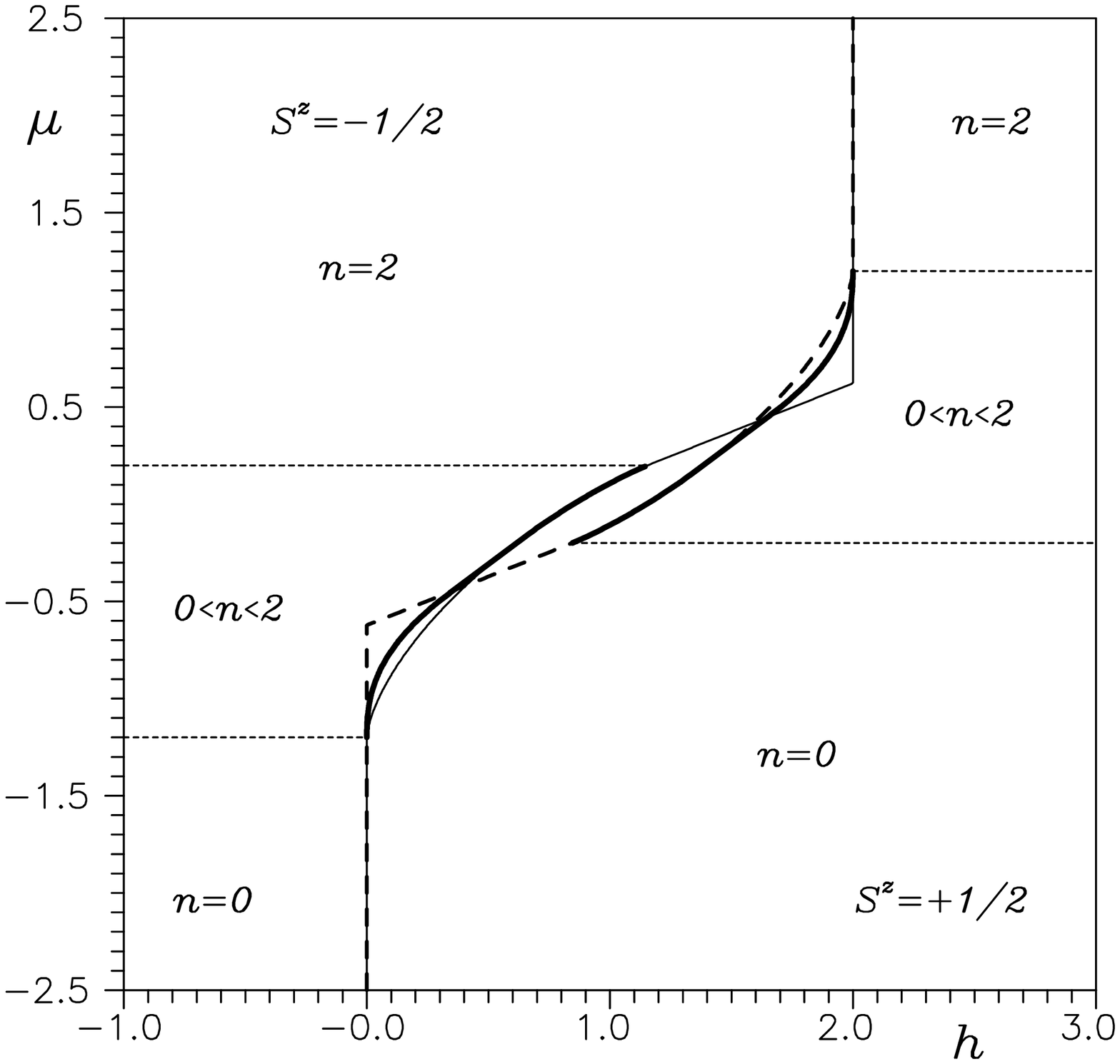}}
  \centerline{b)}
  \caption{Phase diagram $\mu-h$. Dashed and thin solid lines
  surround regions with $S^z=\pm\frac12$, respectively. Thick
  solid line indicate the first order phase transition points.
  a) $g=1$, $W=0.2$; b) $g=1$, $W=0.7$.}
  \label{mu-h}
  \end{figure}

\begin{figure}
  \centerline{\includegraphics[width=0.9\columnwidth]{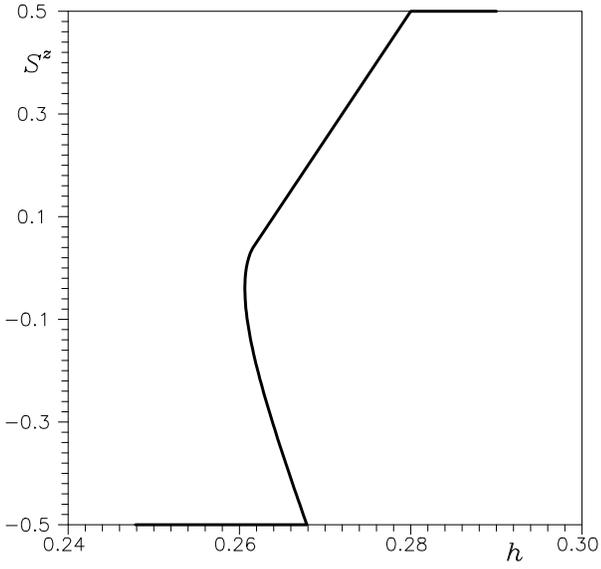}}
  \centerline{a)}
  \bigskip
  \centerline{\includegraphics[width=0.9\columnwidth]{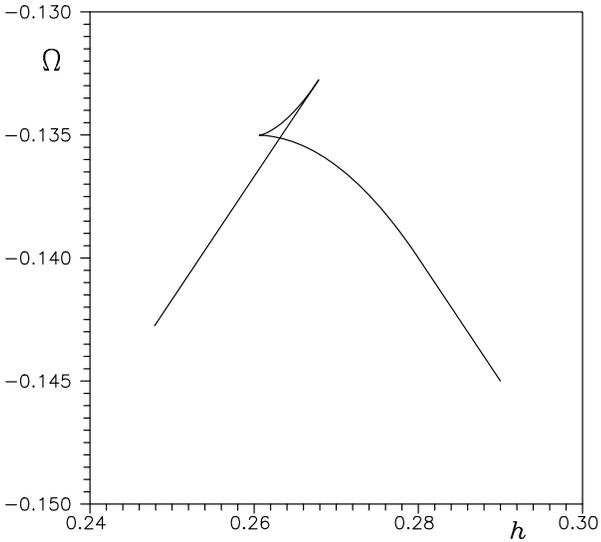}}
  \centerline{b)}
  \caption{Field dependencies of $\left\langle S_z\right\rangle$ (a)
  and thermodynamical potential (b) for $\mu=\mathrm{const}$ regime
  when chemical potential is placed in the lower subband $\mu=-0.37$
  ($W=0.2$, $g=1$, $T=0$).}
  \label{muconst}
  \end{figure}

Integrals in Eqs.~(\ref{partfunc}) and (\ref{thermpot}) can be
calculated analytically for states with $\langle
S^z\rangle=\pm\frac12$ at zero temperature and corresponding phase
diagrams $\mu-h$ which indicate stability regions for these states
are shown in Fig.~\ref{mu-h}a and b for $g>W$ and $g<W$,
respectively.
One can see two regions of $\mu$ and $h$ values where the states
with $\langle S^z\rangle=\pm\frac12$ coexists. In the vicinity of
these regions the phase transitions of first order with the change
of the longitudinal field $h$ and/or chemical potential $\mu$ take
place (see Fig.~\ref{muconst}) and they are shown by thick lines on
phase diagrams (Fig.~\ref{mu-h}).

There are no any specific behaviour when chemical potential is
placed out of bands. If chemical potential is placed in upper
subband the graphs presented in Fig.~\ref{muconst} transform
according to the internal symmetry of the Hamiltonian:
  \begin{equation}
  \mu\to-\mu,\quad h\to 2g-h,\quad n\to 2-n,\quad S^z\to-S^z.
  \end{equation}

\begin{figure}
  \centerline{
  \includegraphics[width=0.9\columnwidth]{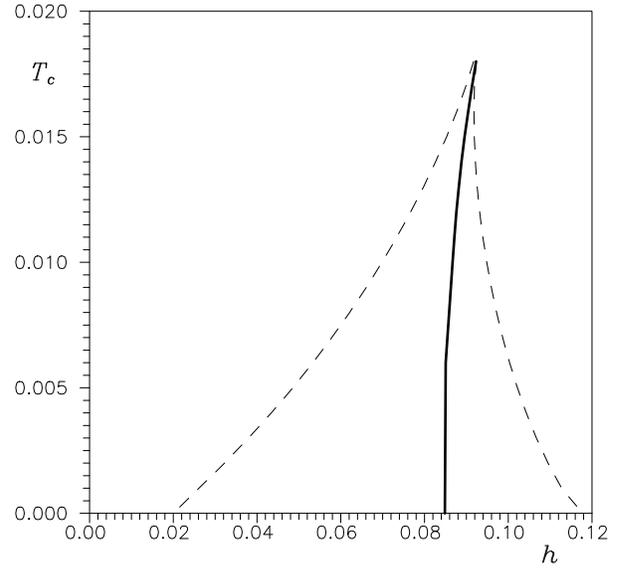}
  }
  \caption{Phase diagram $T_c-h$: solid
  and dashed lines indicate the first order phase transition line
  and boundaries of the phase stability, respectively
  ($g=1$, $W=0.2$, $\mu=-0.5$)}
  \label{ph-mu}
  \end{figure}

With the temperature increase the region of the phase coexistence
narrows and the corresponding phase diagram $T_c-h$ is shown in
Fig.~\ref{ph-mu}. One can see that with respect to Ising model
the phase coexistence curve is shifted in field and distorted from
the vertical line and hence the possibility of the first order
phase transition with the temperature change exists in
pseudospin--electron model for the narrow range of $h$ values.

As it was mentioned above, the band structure is determined by the
pseudospin mean value and its change is accompanied by the
corresponding changes of the electron concentration and for the
$(\mu,h)$ values fixed on the first order phase transition line
there are three solutions for electron concentration one of which
is unstable.

\begin{figure}
  \centerline{
  \includegraphics[width=0.9\columnwidth]{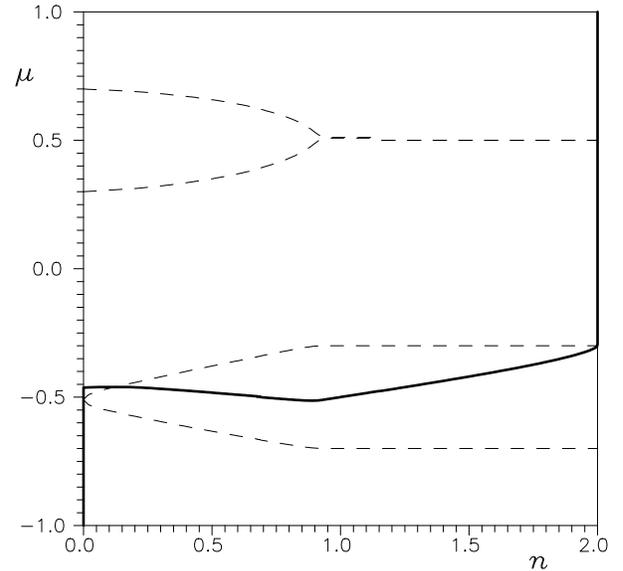}
  }
  \caption{Dependence of the chemical potential $\mu$ and electron
  bands boundaries (dashed lines) on the
  electron concentration $n$ ($T=0.001$, $g=1$, $W=0.2$,
  $h=0.1$).}
  \label{ndepT}
  \end{figure}

\begin{figure}
  \centerline{
  \includegraphics[width=0.9\columnwidth]{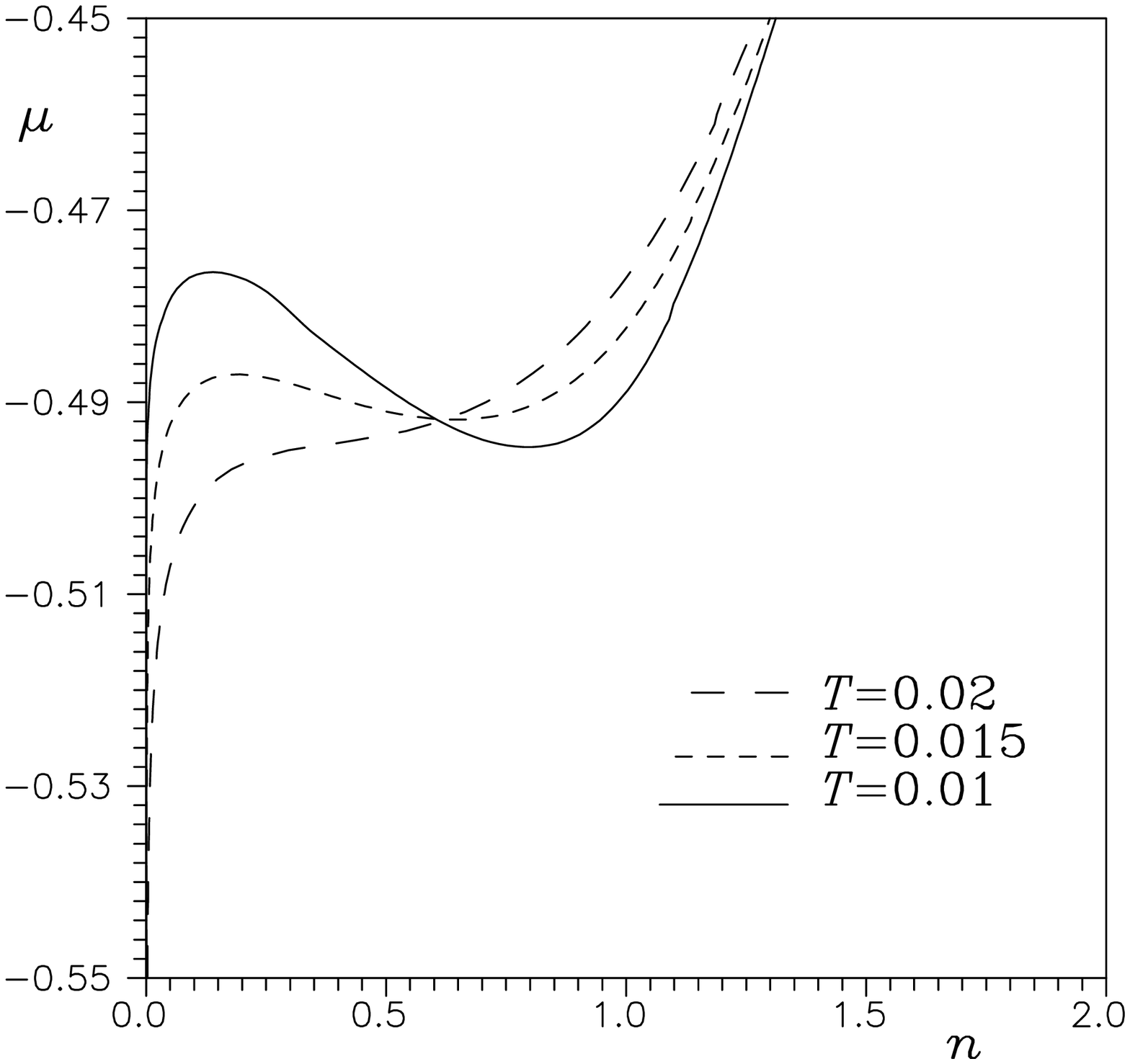}
  }
  \centerline{a)}
  \bigskip
  \centerline{
  \includegraphics[width=0.9\columnwidth]{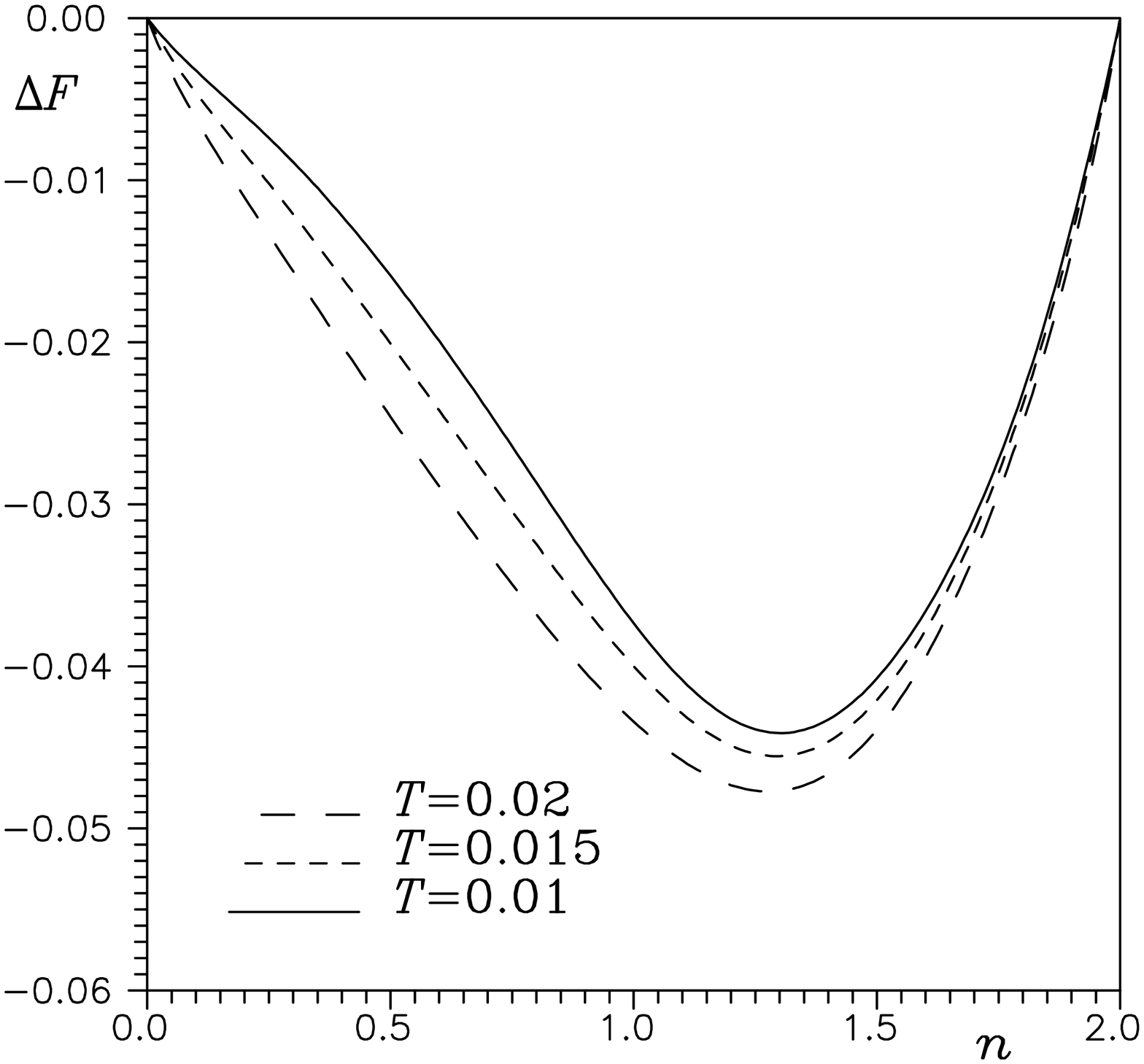}
  }
  \centerline{b)}
  \caption{Dependence of the chemical potential $\mu$ (a) and
  deviation of free energy from linear dependence
  $\Delta F=F(n)-\frac n2 F(2)-\left(1-\frac n2\right) F(0)$ (b)
  on the electron concentration $n$
  for different temperatures~$T$ ($g=1$, $W=0.2$,
  $h=0.1$).}
  \label{ndepT2}
  \end{figure}

In the case of the fixed value of the electron concentration value
(regime $n=\mathrm{const}$) this first order phase transition
transforms into the phase separation. One can see regions with
$d\mu/dn\leq0$, which correspond to this effect in electron
subsystem, on the concentration dependencies (Figs.~\ref{ndepT} and
\ref{ndepT2}a).

\begin{figure}
  \noindent\null\hfill
  \includegraphics[width=0.9\columnwidth]{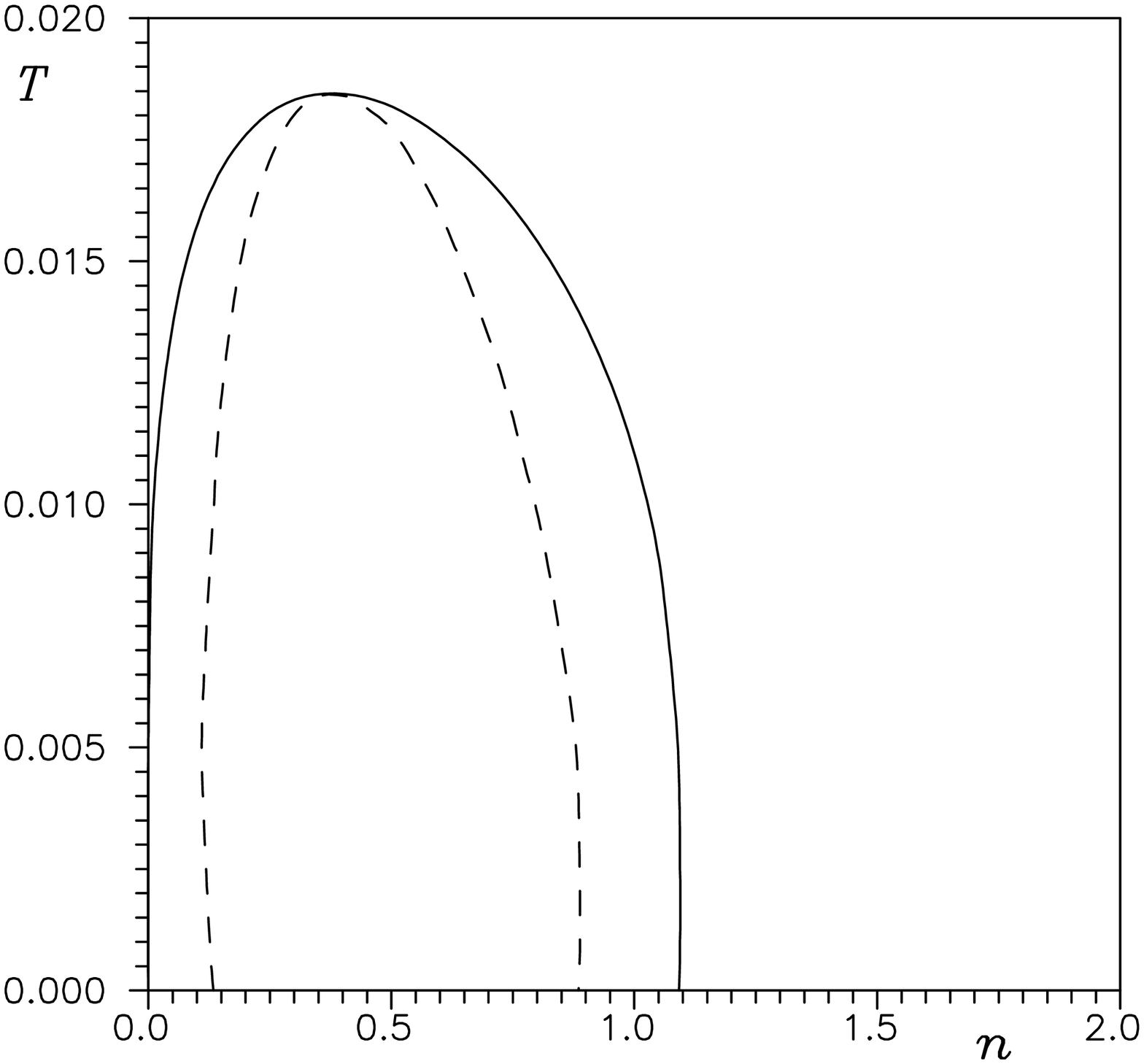}
  \hfill\null
  \caption{Phase diagram $T-n$ for phase separated state:
  solid line --- binodal, dashed line --- spinodal
  ($g=1$, $W=0.2$, $h=0.1$).}
  \label{phdiag}
  \end{figure}

The corresponding dependencies of free energy $F=\Omega+\mu n$ are
given in Fig.~\ref{ndepT2}b. In the phase separated region free
energy deflects up and concentration values at binodal points are
determined by the tangent line touch points or from the chemical
potential dependencies (Fig.~\ref{ndepT2}a) using Maxwell
construction. Resulting phase diagram $T-n$ is shown in
Fig.~\ref{phdiag}.

For the first time the possibility of phase separation in
pseudospin--electron model was marked in \cite{CzJPh} where it was
obtained within GRPA in the limit of strong correlation
$U\to\infty$. Here it is observed for the opposite case of $U=0$.

The problem of phase separation in strongly correlated systems is
not new (see~\cite{Dagotto} and references therein). It was shown
for Hubbard and $t-J$ models \cite{Emery} that for some parameter
values system separates into hole--rich and hole--poor regions with
paramagnetic and antiferromagnetic orders, respectively, and
long--range interaction between these charged regions is considered
as an origin of the appearance of stripe structures. In Ref.
\cite{Freericks} the phase segregation for some parameter values
was reported for the annealed binary alloy with diagonal disorder described
by Falicov--Kimball model. In our case of pseudospin--electron model without
electron correlations system separates into regions with different values of
electron concentration and pseudospin mean value and electron spectrum
contains both wide empty electron band and occupied localized
states of the regions with $n\sim0$ as well as partially filled
wide electron band and empty localized states of the regions with
$n\sim1$ (see Fig.~\ref{ndepT}) the weights of which are determined
by the electron concentration. Such type localized states
(polarons) results from the strong electron--out of
plane apical oxygen vibrations coupling ($g>W$)
in the case of YBaCuO--type structures and it is supposed
that the hopping between them gives
significant contribution in the carrier relaxation observed by the resonant
Raman spectroscopy \cite{Mertelj}.

It should be noted that in the case of spinless fermions
Hamiltonian (\ref{Hamilton}) can be applied for the description of
the oxygen vacancies subsystem in high--$T_c$ superconductors,
which can be treated as quasiequilibrium, and it is known that
their interaction with some relaxation type lattice mode leads to
the phase separation and appearance of superstructures and stripes
\cite{Aligia}.

In this paper we investigated the possible phase transitions in
pseudospin--electron model within DMFT without creation of super
structures ($\vec k=0$) and the phase diagrams presented in
Figs.~\ref{ph-mu} and \ref{phdiag} concern only this case.
In order to detect instabilities associated with a specific wave
vectors one should calculate response functions which will be the
subject of the further investigations.

\begin{acknowledgement}
This work was partially supported by the Ministry of Ukraine for Science and
Technology (project No 2.4/171).
\end{acknowledgement}

\end{document}